# On the time-optimal implementation of quantum Fourier transform for qudits represented by quadrupole nucleus

V.P. Shauro*, V.E. Zobov

L. V. Kirensky Institute of Physics, Siberian Branch of Russian Academy of Sciences,
Academgorodok 50, Krasnoyarsk, Russia

## ABSTRACT

We consider the problem of time-optimal realization of the quantum Fourier transform gate for a single qudit with number of levels $d$ from 3 to 8. As a qudit the quadrupole nucleus with spin $I > \frac{1}{2}$ controlled by NMR is considered. We calculate the dependencies of the gate error on the duration of radio-frequency pulse obtained by numerical optimization using Krotov-based algorithm. It is shown that the dependences of minimum time of QFT gate implementation on qudit dimension are different for integer and half-integer spins.

## 1. INTRODUCTION

One of the important tasks of quantum computer development is the realization of time-optimal gates. It is well known that the duration of gates should be as short as possible in order to minimize the relaxation effects. In most cases there are fundamental limits on the minimum time of gate implementation, which depends on the Hamiltonian of the quantum system and type of the gate required[1]. If the time of gate implementation is less than minimal time, the gate always has a finite error even in case of absence of interaction with the environment. In terms of quantum optimal control theory the task of finding the time-optimal quantum gates is as follows[1]. Suppose there is a closed quantum system that is governed by the Schrödinger equation with the Hamiltonian

$$H(t) = H_0 + \sum_k u_k(t) H_k \ . \tag{1}$$

Here $H_0$ is the time-independent drift Hamiltonian and $H_k$ are the terms describing the interactions with the external control fields which has time-dependent amplitudes $u_k(t)$. We should find the functions $u_k(t)$ that the evolution operator of a quantum system,

$$U(T) = \hat{D} \exp\left( -i \int_0^T H(t) dt \right), \tag{2}$$

performs the desired logical transform $U_f$ (quantum gate). Here $\hat{D}$ is the time-ordering operator. In finding the $u_k(t)$ we should aim to the minimum value of the time $T$, while the gate error,

$$\Delta = 1 - \left| Tr\left( U_f^+ U(T) \right) \right|^2 / Tr^2(\mathbf{1}) \ , \tag{3}$$

remains below the error threshold for fault-tolerant quantum computing[2,3]. This problem can be solved analytically only in some special cases for one or two qubits[1] (e.g. ½-spins). In most cases the various numerical methods are used for the search of functions $u_k(t)$ such as GRAPE[4,5] and Krotov-based[5-8] algorithms.

*rsa@iph.krasn.ru

One of the important gates in quantum computing is a quantum Fourier transform[3, 9] (QFT). The matrix representation of this gate in general case of $d$-level quantum system has the form

$$QFT_d = \frac{1}{\sqrt{d}} \begin{bmatrix} 1 & 1 & 1 & \cdots & 1 \\ 1 & \delta & \delta^2 & \cdots & \delta^{d-1} \\ 1 & \delta^2 & \delta^4 & \cdots & \delta^{2(d-1)} \\ \vdots & \vdots & \vdots & \ddots & \vdots \\ 1 & \delta^{d-1} & \delta^{2(d-1)} & \cdots & \delta^{(d-1)^2} \end{bmatrix}, \quad \delta = \exp\left(\frac{2\pi i}{d}\right). \tag{4}$$

QFT is widely used in many quantum algorithms. The most striking example is Shor's algorithm[3, 9], where the use of QFT allows to solve the problem of factorization in polynomial number of operations.

The evaluation of minimum time of QFT gate realization was obtained by Schulte-Herbrüggen et al.[10] for linear chain of $n$ ½-spins (qubits) each of which is controlled separately by resonance radio-frequency (RF) field, that is by $n$ RF fields. The several ways of implementation has been considered including the numerical search of the optimized pulse by the GRAPE algorithm. Some main results are follows. First, the dependence of minimum time on number of qubits remains the same qualitatively (namely, linear on $n$ or logarithmic on $d=2^n$) for both the control schemes with decomposition of QFT gate on simplest gates and the simultaneous control of qubits by optimized RF pulses. Second, the time of QFT gate implementation can be significantly reduced in the latter case.

The QFT gate can be realized in more complex case of multilevel ($d$-level) quantum systems called qudits, e.g. multilevel atoms[11] or quadrupole nuclei[12]. The advantage is the reduction of number of quantum systems to $\log_2 d$ times compared to the binary case for the same dimension of Hilbert space. However for a single qudit the time of control increased. This time depends significantly on physical system under consideration. In this paper, we investigate the dependence of the minimum time of QFT gate implementation on the number of levels $d = 2I + 1$ for quadrupolar nuclei with spin $I > ½$ controlled by NMR.

## 2. OPTIMAL CONTROL OF QUADRUPOLE NUCLEUS

The convenient physical model, which can be considered as a qudit, is quadrupole nucleus controlled by NMR. The Hamiltonian of a nucleus with spin $I > ½$ in the reference frame rotating with the frequency of external control RF field is[13]

$$H = (\omega_{rf} - \omega_0)I_z + q(I_z^2 - \tfrac{1}{3}I(I+1)) + u_x(t)I_x + u_y(t)I_y \tag{5}$$

The first Zeeman term vanishes, since we assume that the frequency of RF field $\omega_{rf}$ is equal to the Larmor frequency $\omega_0$. The second term is the quadrupole interaction of nucleus with the gradient of crystal field, where $q$ is the constant of this interaction. Here $I_\alpha$ is the operator of spin projection on the α axis and the functions $u_\alpha(t)$ are the projections of the control field on the corresponding axes (for brevity, we call it as amplitude of the field). In the absence of RF field, the system has $d = 2I+1$ nonequidistant energy levels corresponding to the states with the different values of spin projection $I_z$. These states are used as the computation basis for a qudit.

As in any quantum system the time scale of quantum operations related to the weakest interaction that leads to nonequidistant spectrum. In our model it is the quadrupole interaction. Therefore, for convenience as a relative time unit we take the reverse unit of constant $q$.

The numerical search of optimal control field that realize the QFT gate carried out using the Krotov-based algorithm. The basic idea underlying the algorithm is to find the maximum of functional

$$J = \left| \left\langle U_f \middle| U(T) \right\rangle \right|^2 - \lambda \int_0^T \sum_k \left( u_k(t) - v_k(t) \right)^2 dt - 2 \operatorname{Im} \int_0^T dt \left\langle B(t) \middle| (i\partial_t - H(t)) \middle| U(t) \right\rangle. \tag{6}$$

In this functional the first term determines the fidelity of the gate. The second one is a limitation on either the field amplitude or the pulse shape. This term can be written in different ways, depending on the context of the problem[8]. Because for our theoretical task it is necessary to exclude additional restrictions on pulse we have used a modified scheme from the work Eitan, Mundt and Tannor[8]. In this case, while the reference function $v(t)$ is chosen as equal to $u(t)$ at the previous step of the algorithm (see below), the limitation on the amplitude goes to zero with approaching to the maximum of functional. The third term with Lagrange multiplier $B(t)$ is necessary to ensure that the solution satisfy the Schrödinger equation. Equating the functional variation to zero, we obtain a system of equations with boundary conditions[6, 7] that used to construct a numerical iterative algorithm after discretization of time interval. The main scheme is as follows[7]:

1. Guess initial controls $u_k(t_n)$, where $t_n = n\Delta t$, $n = 1, ..., N$ and $\Delta t = T/N$ is the discrete time step;

2. Starting from $U(0)=E$ ($E$ is a unit matrix), calculate the evolution $U(t_n)=U(t_{n-1})U(t_{n-2})...U(t_1)U(0)$ for all $t_n$;

3. Starting from $B(T) = U_f \left\langle U_f \big| U(T) \right\rangle$, calculate the "reverse" evolution $B(t_{n-1})=B^*(t_n)B^*(t_{n+1})...B^*(t_{N-1})B(T)$ for all $t_n$;

4. Using the equation $u_{km}(t_n) = u_{k(m-1)}(t_n) - \frac{1}{\lambda}\text{Im}\left\langle B_{m-1}(t_n) \big| H_k(t) \big| U_m(t_n) \right\rangle$, update the amplitude at all points consistently updating the evolution operator $U_m(t_n)$ for all $t_n$ (forward propagation), where $m$ is the iteration number;

5. Using the equation $\tilde{u}_{km}(t_n) = \tilde{u}_{k(m-1)}(t_n) - \frac{1}{\lambda}\text{Im}\left\langle B_m(t_n) \big| H_k(t) \big| U_m(t_n) \right\rangle$, update the amplitude at all points consistently updating operator $B_m(t_n)$ for all $t_n$ (backward propagation);

6. Repeat steps 4-5 before reaching stopping criterion $\Delta_m - \Delta_{m+1} < \varepsilon$ with $\varepsilon \sim 10^{-10}$.

The parameter $\lambda$ depends mainly on the step $\Delta t$ and it is chosen empirically. The choice of its value defines the rate of convergence and the numerical stability of the algorithm[7]. With the each cycle (steps 4-5) the pulse shape changes gradually reducing the error of gate desired. A more detailed description of the algorithm can be found in references[6-8].

In our calculations for $U_f = QFT_d$ (4) we set $N$ from 100 to 200, depending on the dimension of qudit and the pulse duration. The parameter $\lambda$ is varied from $100\Delta t$ to $500\Delta t$ correspondingly. As an initial guess we take the random values of pulse amplitude at every ten points with subsequent interpolating at the remaining points by cubic spline. The Figure 1 shows the example of optimized pulse in simplest case of spin $I$=1 ($d$=3, qutrit). There are much more sophisticated pulse shapes for $d > 3$.

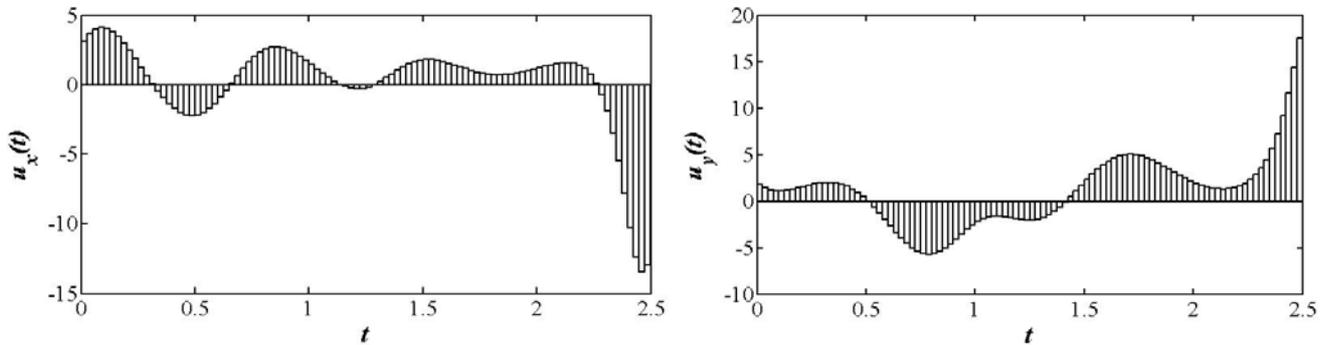

Figure 1. Optimized pulse for QFT gate realization in case of spin $I$=1 at $T$=2.5/$q$ and $N$=100. The gate error in simulation is $\Delta < 10^{-8}$.

## 3. RESULTS

Having calculated optimized pulses to implement the QFT gate on system (5) with different $T$, we can determine the minimum time, $T_{min}$, as the time at which the gate error is below the threshold value, such as $\Delta < 10^{-5}$.

However, we must note one issue which was pointed out by Schulte-Herbrüggen et al.[10]. The quantum gates belong to the group of unitary operators, while the evolution operator belongs to a group of special unitary operators. Thus, we can obtain the desired gate only up to a global phase factor

$$U(T) = e^{i\varphi}U_f, \quad \varphi = \varphi_0 + \frac{2\pi k}{d}, \quad k = 0,1,...,d-1 \tag{7}$$

Here $\varphi_0$ is determines from conditions $\det\left(e^{i\varphi_0}U_f\right) = +1$, $\varphi_0 \in [0,\pi]$. The difficulty is that the definition of gate with different global phase, $\varphi$, can leads to different minimum time of the gate implementation. This has been demonstrated by numerical simulations for QFT gate on a linear chain of three ½-spins[10] and for QFT and SWAP gates on two ½-spins[14].

The Figure 2 shows the results of our calculation for QFT gate implementation on spin $I$ = 1. For the three possible phase factors we obtain very different values of the minimum time. Note that the solution with $\varphi$ =9π/6 was obtained previously[12]. In the case of spin $I$=3/2 (4-level qudit) we have already four curves (Figure 3), two of which are almost coincided.

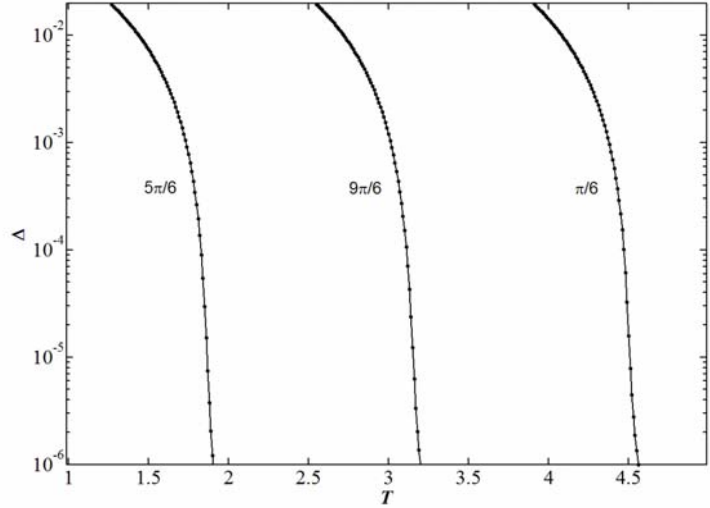

Figure 2. The dependence of the QFT gate error on the duration of optimized pulse for spin $I$ = 1 ($d$ = 3) with different values of the global phase $\varphi$ (denoted beside the curves)

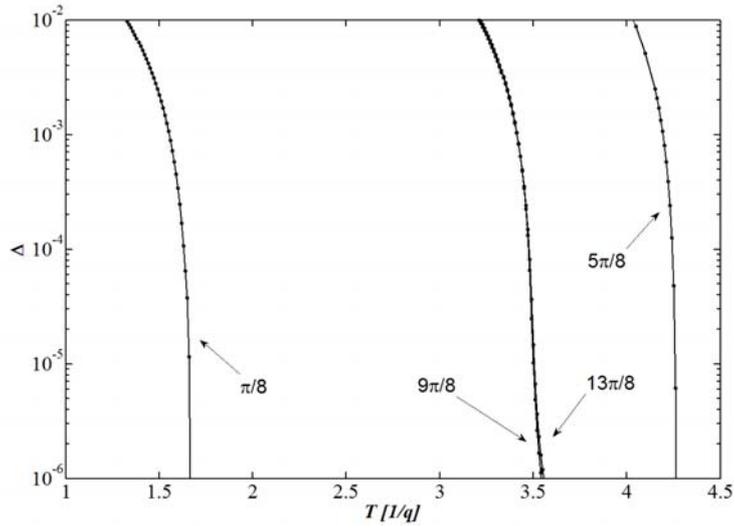

Figure 3. The dependence of the QFT gate error on the duration of optimized pulse for spin $I$ = 3/2 ($d$ = 4) with different values of the global phase $\varphi$ (denoted beside the curves)

The calculation of the gate error dependence on the pulse duration for each value of the global phase is very time-consuming for large values of $d$. At the same time we do not need to obtain solutions for the all possible phase factors, but only the solution with minimum time is needed. Therefore, for $d > 4$, we have used the approach of Schulte-Herbrüggen et al.[10] Namely, for a fixed time $T$ the several tens runs was performed with 10000 iterations and with different initial guess $u_k(t)$. From this solutions we chosen the one with minimum error and perform further calculation until the error is almost unchanged from iteration to iteration. Follow this procedure for several values of the time $T$, we can roughly estimate the required minimum time. To more accurate estimate the calculations was performed by the PFT method[14]. Its essence is that given the solution with small gate error for the time $T$, we perform the calculation for the time $T-\Delta T$ setting the pulse generated for time $T$ as initial guess. As a result, the computation time is significantly reduced for obtaining dependencies as in Figure 2-3. By these methods, we have performed the calculations for the values of $d$ from 5 to 8. The result is shown in Figure 4.

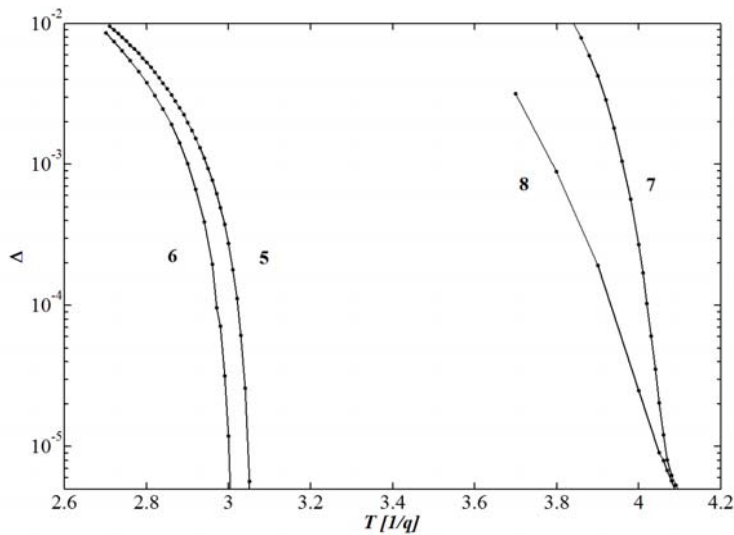

Figure 4. . The dependence of the QFT gate error on the duration of optimized pulse for $d$=5, 6, 7, 8 (denoted by numbers beside the curves)

Now we can plot the dependence of minimum time on number of levels $d$ by determining the time at gate error $\Delta=10^{-5}$. The result is shown in Figure 5. The interesting result is that the obtained points lie on a smooth line only if ones are drawn for odd and even $d$ separately. Probably the difference is due to some features in the structure of the Hamiltonian for integer and half-integer spins. For example, the off-diagonal matrix elements of Hamiltonian (5), which coupling the central diagonal matrix elements of the integer spins in contrast to half-integers, had a significant impact on the structure of the sequence of non-selective pulses to implement the selective rotation gate on qudits[15]. Similar differences can also affect in numerical optimization.

## 4. CONCLUSIONS

As can be seen from the figures, the realization times of QFT gate with d = 3 and 4 are very close, as well as for d = 5, 6 and d= 7, 8. This result is interesting because it allows to perform a quantum computation with larger state basis but for the same time. Contrary to the case of multi-field control of qubits, in our model we use the single RF field. It leads to more difficulties in the calculations because of very complicated pulse shape and extremely low convergence in most cases. Presently available data are not sufficient for reliable determination of dependency minimum time on number of levels d. However, presented results can be useful in order to estimate a minimum time of QFT gate for systems of many quadrupole nuclei.

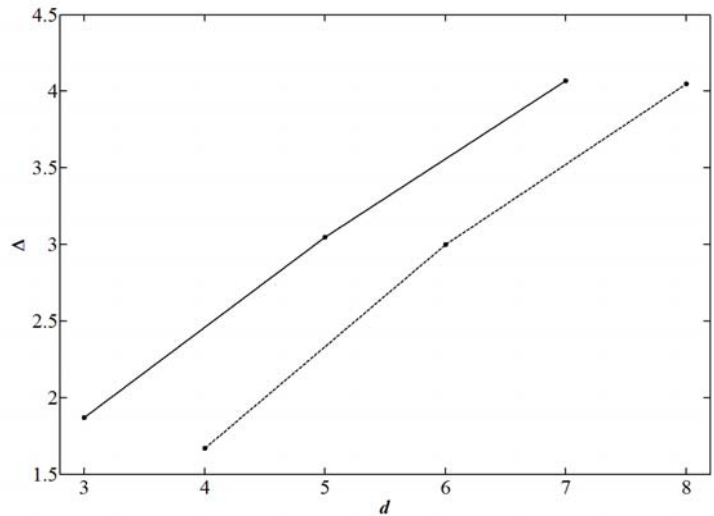

Figure 5. The dependence of the minimum time of QFT gate realization on the number of levels $d$ of qudit.